\documentclass[conference]{IEEEtran}
\usepackage{array}
\usepackage{multirow}% http://fctan.org/pkg/multirow

\usepackage[noadjust]{cite}
\usepackage{algorithm}
\usepackage{algorithmicx}
\usepackage{algpseudocode}
\usepackage{amssymb}
\usepackage{bm}
\usepackage{amsmath}
\usepackage{cases} 
\usepackage{multirow}
\usepackage{booktabs}
\usepackage[caption=false,font=footnotesize]{subfig}
\usepackage{siunitx}
\usepackage{tabularx}
\usepackage{makecell}
\usepackage[table,xcdraw]{xcolor}
\usepackage{booktabs}

\usepackage{setspace}
\ifCLASSINFOpdf
  \usepackage[pdftex]{graphicx}
  \graphicspath{{./}{./}}
  \DeclareGraphicsExtensions{.pdf,.jpeg,.png}
\else
  \usepackage[dvips]{graphicx}
  \graphicspath{{./}}
  \DeclareGraphicsExtensions{.eps}
\fi
\begin{document}

\title{
A Constituent Codes Oriented Code Construction Scheme for Polar Code-Aim to Reduce the Decoding Latency
\vspace{-0.5em}}

\author{\IEEEauthorblockN{Tiben Che and Gwan Choi}
\IEEEauthorblockA{Department of Electrical and Computer Engineering\\
 Texas A\&M University, College Station, Texas 77840\\
 Email: $\lbrace$ctb47321, gwanchoi$\rbrace$@tamu.edu}
}


%Telephone: (800) 555--1212, Fax: (888) 555--1212}


\maketitle

\begin{abstract}
%\boldmath
This paper proposes a polar code construction scheme that reduces constituent-code related decoding latency.
Constituent codes are the sub-codewords with specific patterns.
They are used to accelerate the successive cancellation decoding process of polar code with negligible performance degradation.
%We explore this idea in this paper to further reduce the decoding latency.
We modify the traditional construction approach to yield increased number of desirable constituent codes that speeds the decoding process.
For $(n,k)$ polar code, instead of directly setting the $k$ best and $(n-k)$ worst bits to the information bits and frozen bits, respectively, we swap the locations of some information and frozen bits carefully according to the qualities of their equivalent channels.  
%The  constituent codes oriented polar code construction algorithm is presented.
We conducted the simulation of $1024$ and $2048$ bits length polar codes with multiple rates and analyzed the decoding latency for various length codes.
The numerical results show that the proposed construction scheme generally is able to achieve at least around $20\%$ latency deduction with an negligible loss in gain with carefully selected optimization threshold.   
%Discussions are also presented.  

\end{abstract}

% \fi
\IEEEpeerreviewmaketitle

\section{Introduction}
\label{Introduction}
Recently, polar code~\cite{arikan2009channel} attract more and more research interests due to that it is the first code which provably achieves the channel capacity and its low coding complexity. 
%Its low-complexity construction and decoding schemes makes it very promising for real application as well.
Such property makes it very promising for real scenario such as wireless communication and storage.
Successive cancellation (SC)~\cite{arikan2009channel}, list successive cancellation (LSC)~\cite{tal2011list} and belief propagation (BP)~\cite{hussami2009performance} are the three most widely known decoding algorithms.  
Among those, SC and LSC receive more attention due to their simpler hardware complexity compared with that of BP. 
%Actually, LSC can be regarded as an extension of SC since they both do have the same decoding core. 
Due to its serial property, SC decoder suffers from the high decoding latency.
LSC considered as an extension of SC, similarly, has the same problem. 
The latency reduction of SC decoder is able to benefit the LSC decoder as well.
Thus, a lot of efforts have been done on the SC decoder to reduce the latency from both hardware and algorithm aspects.
% Now mainly there are three kinds of algorithm for polar codes decoding.
% E. Arikan in~\cite{arikan2009channel} presents a successive cancellation (SC) algorithm which is able to successively accomplish decoding with recursive cancellation. 
% I. Tal~\cite{tal2011list} makes the SC algorithm more competitive by exploring more paths among the codewrds tree; and this method is referred as list successive cancellation (LSC).    
% Also, N. Hussami et al. in~\cite{hussami2009performance} shows that the belief propagation (BP) can be applied as decoding algorithm.

C. Leroux~\cite{leroux2011hardware} proposed both tree and line architecture of SC decoder. For length $n$ polar code, it takes $(2n-2)$ clock cycles to decode. 
Later on, he~\cite{leroux2013semi} proposed a semi-parallel architecture for both tree and line SC decoder, which makes a trade-off between hardware complexity and latency.
C. Zhang~\cite{zhang2013low} proposed a low latency SC decoder with pre-computing and overlapped architecture. 
Pre-computing technology reduces the latency to $(n-1)$ clock cycles, and the overlap scheme significantly the throughput with multiple frame situation.
B. Yuan~\cite{yuan2014low} proposed an architecture with applying the 2-bit decoding at the last stage and some gate level optimizations, this further reduce the latency to $(3/4n-1)$ clock cycles.
Alamdar-Yazdi~\cite{alamdar2011simplified} proposed the simplified SC (SSC)  which can significantly reduce the latency via some certain pattern sub-codewords. 
These kinds of sub-codewords are also called constituent codes.
This is the first time the concept of constituent codes has been mentioned.
Inspired by this, Sarkis~\cite{sarkis2014fast} proposed the fast-SSC which can further reduce the latency by exploring more kinds of constituent codes. 
T. Che~\cite{che2016tc} proposed the hardware architecture of constituent codes based polar codes decoder. It allows the decoding processes are compatible with both conventional and constituent codes based polar codes.
P. Giard~\cite{giard2014237} proposed an unrolled architecture with fast-SSC is able to achieve 237 Gbps throughput. 
P. Giard~\cite{giard2016fast} also proposed a low-complexity decoder for low rate polar code. In that work, he further utilized the potential of constituent codes by changing the frozen and information sets. 
Additionally, the idea of of constituent codes also benefits other decoding algorithm.
J. Xu~\cite{xu2015xj} applied the constituent concept to the BP decoding which significantly reduces the computing complexity.  
G. Sarkis~\cite{sarkis2016fast} proposed an constituent codes based LSC decoding. 
T. Che~\cite{che2015overlapped} also pointed that the constituent codes can benefit the overlapped LSC architecture in term of hardware efficiency.

Introducing constituent code is an approach to reduce decoding latency.
Most of the aforementioned works stress on the decoding sides.
In this paper, we explore the potential of constituent codes from an opposite angle.
We stress on the construction scheme to make the codeword more constituent-code-friendly.
By adjusting the traditional construction approach, more expected types of constituent codes are manually produced for decoding.
For $(n,k)$ polar code, instead of directly setting the $k$ best and $(n-k)$ worst bits to the information bits and frozen bits, respectively, we thoughtfully swap the locations of some information and frozen bits according to the qualities of their equivalent channels.  
The constituent codes oriented polar code construction algorithm is described.
%The corresponding algorithm to optimize the constituent codes division is also described.
We conducted the simulation of $1024$ and $2048$ length polar codes with multiple rates and analyzed the decoding latency for various length codes. 
The numerical result shows that the proposed construction scheme typically achieves 4-20$\%$ latency reduction with negligible loss in decoding performance with carefully selected optimization threshold.  
Some relevant discussions are also presented.

This paper is organized as follows. 
The relative background is reviewed in section~\ref{Background}. 
Then, the proposed construction scheme are described in section~\ref{Proposed construction Scheme}. 
After that, the numerical results and relevant discussions are presented in section~\ref{Simulation Results and Discussions}. 
Finally, this paper is concluded in section~\ref{Conclusion}.

\section{Background}
\label{Background}
\subsection{Polar code}
\label{Polar Code}

\begin{figure}[!t]
\centering
\subfloat[]{\includegraphics[width=1.6in]{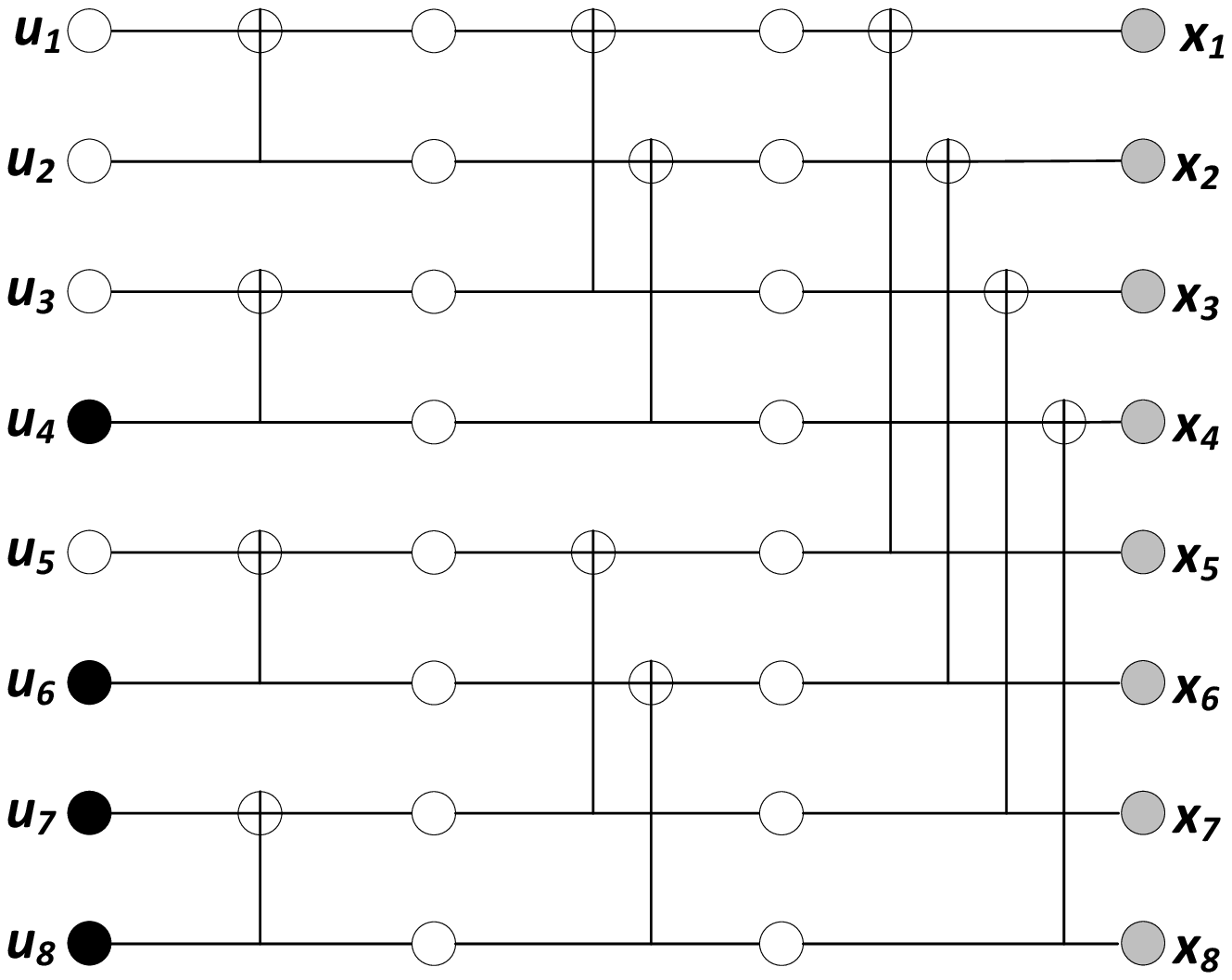}\label{encoder}}
\hfil
\subfloat[]{\includegraphics[width=0.8in]{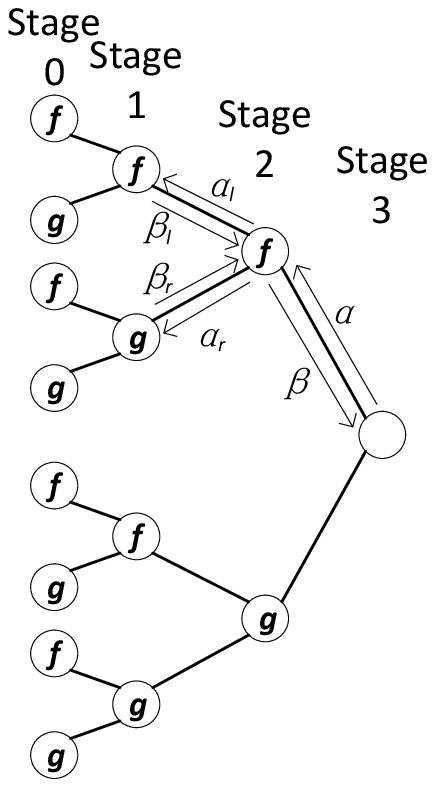}\label{SC_tree}}
\hfil
\caption{\protect\subref{encoder} Encoder of $(8,4)$ polar code, 
\protect\subref{SC_tree} Tree presentation of $(8,4)$ SC decoder
}
\label{polar_code_introduction}
\end{figure}

As described by E. Arikan~\cite{arikan2009channel}, a polar code is constructed by successively performing channel polarization. 
Polar codes are linear block codes of length $n = 2^m$. 
The coded codeword ${\bm{x}}\triangleq {(x_1,x_2,\cdots,x_n)}$ is computed by $\bm{x}=\bm{u}\bm{G}$ where $\bm{G=F^{\otimes m}}$, and $\bm{F^{\otimes m}}$ is the $m$-th Kronecker power of 
$\bm{F} = 
\begin{bmatrix}
1&0\\
1&1
\end{bmatrix}
$. 
Each row of $G$ corresponds to an equivalent polarizing channel. 
For an $(n,k)$ polar code, $k$ bits that carry source information are called information bits. 
They are transmitted via the $k$ best channels. 
While the rest $n-k$ bits, called frozen bits, are set to zeros and are placed at the $n-k$ worst channels. 
Fig.~\ref{encoder} shows an example of the construction of 8-bit polar code, where information bits and frozen bits are denoted by black nodes and white nodes on the most left side, respectively.

Polar codes can be decoded by recursively applying successive cancellation to estimate $\hat{u}_i$ using the channel output $y_{0}^{n-1}$ and the previously estimated bits $\hat{u}_{0}^{i-1}$. 
This method is referred as successive cancellation (SC) decoding.
Actually, SC decoding can be regarded as a binary tree traversal as described in Fig.~\ref{SC_tree}.
The number of bits of one node in stage $m (m = 0,1,2...)$ is equal to $2^m$. 
$\bm{\alpha}$ stands for the soft reliability value, typically is log-likelihood ratio (LLR). 
Each left and right child nodes can calculate the LLR for current node via $f$ and $g$ functions, respectively~\cite{leroux2013semi}. 
However, in order to compute $g$ function, a feedback $\bm{{\beta}_l}$ from left child of the same parent node is needed. This kind of feedback is called partial sum.
Actually, this serial property of feedback operation limits the throughput of SC decoding.

%Determining the location of the information and frozen bits depends on the channel model and the channel quality is investigated in~\cite{tal2013construct}.
%\begin{figure}[!tc]
%\centering
%\includegraphics[width=3in]{code_tree.eps}
%\caption{An example of LSC decoding with list size 4 for $(8,4)$ polar code from codeword tree aspect}
%\label{code_tree}
%\end{figure} 

\subsection{Constituent codes based SC decoding}
\label{Constituent codes based SC decoding}

%SC decoding generally suffers from the high latency due to its inherent serial property.
The recursive processing of getting the partial sum from each node significantly constrains the decoding speed.
Thus, in order to obtain partial sum directly without performing tree traversal, constituent codes based SC decoding has been proposed~\cite{alamdar2011simplified}, \cite{sarkis2014fast}. 
Certain patterns in the codewords allows us to decode the sub-codewords and get their corresponding partial sums immediately, which significantly reduces the partial-sum-constrained latency.

\begin{figure}[t]
\centering
\subfloat[]{\includegraphics[width=1.2in]{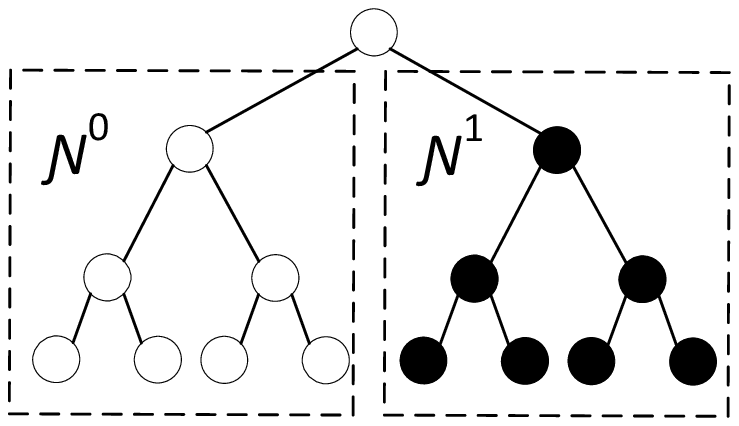}\label{n0n1}}
\hfil
\subfloat[]{\includegraphics[width=1.2in]{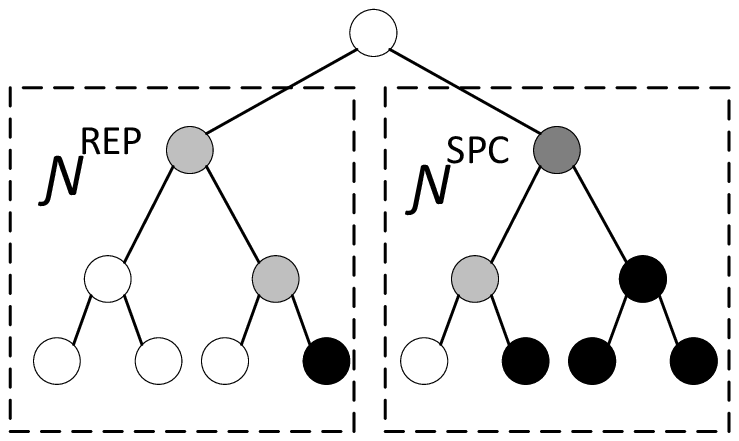}\label{nsnr}}
\caption{SC decoding tree simplified by constituent codes
%\protect\subref{n0n1} An example of $\mathcal{N}^0$ and $\mathcal{N}^1$ in a 8-bit %polar code tree, and \protect\subref{nsnr} An example of $\mathcal{N}^{SPC}$ and $%\mathcal{N}^{REP}$ in a 8-bit polar code tree
} 
\label{2tree}
\end{figure}

$\mathcal{N}^0$, $\mathcal{N}^1$, $\mathcal{N}^{SPC}$ and $\mathcal{N}^{REP}$ are the four most common constituent codes.
$\mathcal{N}^0$ and $\mathcal{N}^1$ only contain either frozen bits or information bits, respectively.
%For $\mathcal{N}^0$ codes, we can set the corresponding to $0$ immediately. 
%For $\mathcal{N}^1$ node, the partial sums can be directly determined via threshold detection Eq.~(\ref{hard_decision}). 
$\mathcal{N}^{SPC}$ and $\mathcal{N}^{REP}$ contain both frozen bits and information bits.
In the $\mathcal{N}^{SPC}$ codes, only the first bit is frozen. It makes the length $n$ constituent codes as a rate $(n-1)/n$ single parity check (SPC) code.
%This code can be decoded by performing parity check with the least reliable bit. Typically it's the one with minimum absolute value of LLR. 
In the $\mathcal{N}^{REP}$ codes, only the last bit is information bit.
In this case, all the corresponding partial sums should be the same since they all are the reflections of the last information bit. 
All the above four constituent codes can be decoded very quickly.
%Thus, the decoding algorithm starts by summing all input LLRs and the partial sums is calculated by performing the hard detection to the final summary. 
According to T. Che's implementation ~\cite{che2016tc}, the latency of length $n$ constituent code can be reduced from $2n-2$ to 1, 1, $log_2n+1$ and $log_2n$ for $\mathcal{N}^0$, $\mathcal{N}^1$, $\mathcal{N}^{SPC}$ and $\mathcal{N}^{REP}$ codes, respectively.
Fig.~\ref{2tree} shows an example of how constituent code can simplify the SC decoding tree.

%\subsection{Current existing partial sums generator}
%\label{Current existing partial sums generator}
%
%\subsubsection{Feed-forward partial sums generator}
%\label{Feed-forward partial sums generator}

\section{Proposed construction Scheme}
\label{Proposed construction Scheme}
%\begin{figure}[!tc]
%\centering
%\includegraphics[width=3in]{proposed_design.eps}
%\caption{The architecture of proposed design}
%\label{proposed_design}
%\end{figure} 
% In this section, we first illustrate the basic ideal of constituent code optimization. 
% Then, the details of the optimization algorithm is presented. 

% \subsection{Constituent Code Optimization }
% \label{Constituent Code Optimization}

As described before, constituent codes are decoded faster than conventional polar codes.
Thus, in order to reduce the decoding latency, we expect more constituents codes, especially constituent codes with large length. 
According to the definition of constituent codes, the initial distribution of constituent codes is determined by the location of information and frozen bits. 
If we can change the locations, we are able to manually produce expected constituent codes.
However, the location of frozen and information bits are very sensitive, and random changes may cause negative influence on the coding performance. 
Thus, a thoughtful construction scheme to produce more expected constituent code is on demand.  
The division of information and frozen bits is decided by the qualities of the equivalent channels which are corresponding to each bit.
Based on the channel model, the equivalent channel qualities can be calculated accordingly~\cite{arikan2009channel}~\cite{tal2013construct}~\cite{trifonov2012efficient}~\cite{vangala2015comparative}. 
This gives us a hint that if we swap some information and frozen bits those with similar channel qualities, this might only incur a very slight performance loss.
The numerical simulation results in the following section prove this idea. 
In this work, binary erasure channel (BEC) is used as our channel model, and thus Bhattacharyya parameter is used as the metric for equivalent channel quality.
This method can be extended to any other kind of channel model. 

Now, we have the idea about how to change the division of constituent codes. 
Next, we need to consider what kind of changes are desired. 
For any length $n$ polar code, it can be regarded as a combination of the following four types of sub-codewords.
%its subset of leaf nodes can compose any of the following four types of sub-tree.
\begin{itemize}
  \item Type-I: All the bits are frozen bits. This is also $\mathcal{N}^0$ constituent code.  
  \item Type-II:All the bits are information bits. This is also $\mathcal{N}^1$ constituent code. 
  \item Type-III: Only one bit is information bit, the rest are frozen bits. This can be regarded as the combination of one $\mathcal{N}^{REP}$ and multiple $\mathcal{N}^0$ constituent codes. 
  \item Type-IV: Only one bit is frozen bit, the rest are information bits. This can be regarded as the combination of one $\mathcal{N}^{SPC}$ and multiple $\mathcal{N}^1$ constituent codes.
\end{itemize}

\begin{figure}[t]
\centering
\subfloat[]{\includegraphics[width=1.6in]{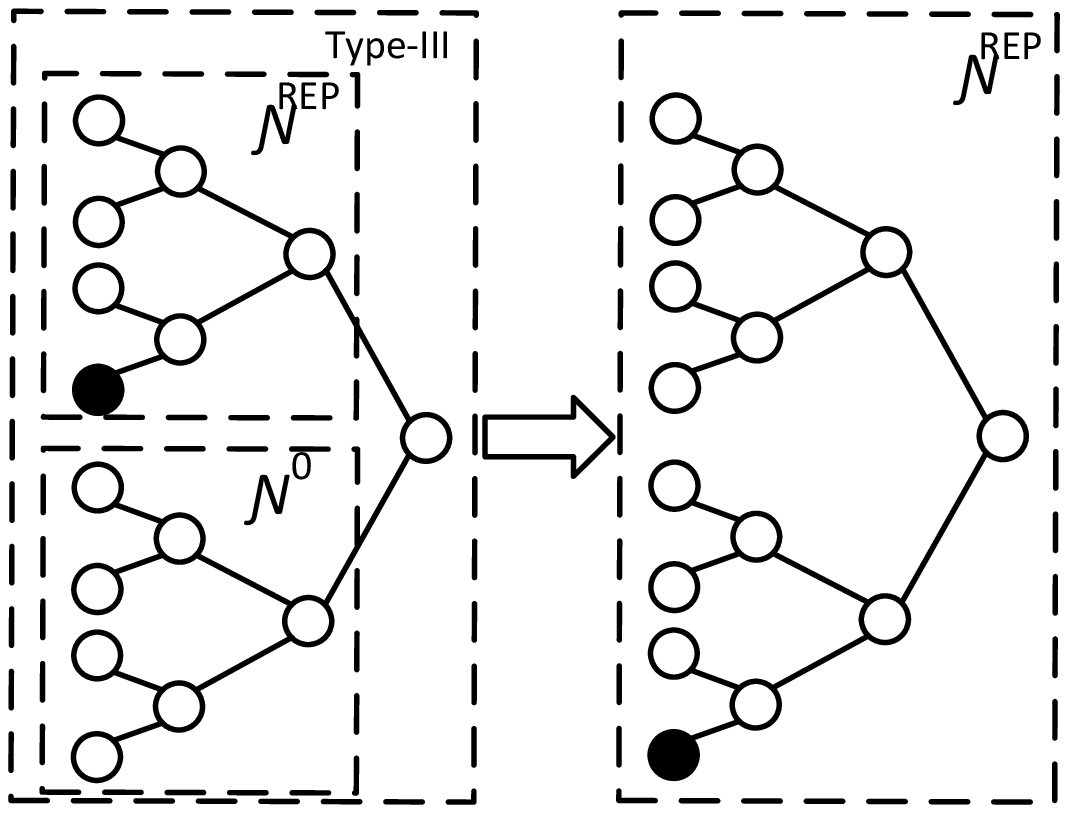}\label{move}}
\hfil
\subfloat[]{\includegraphics[width=1.6in]{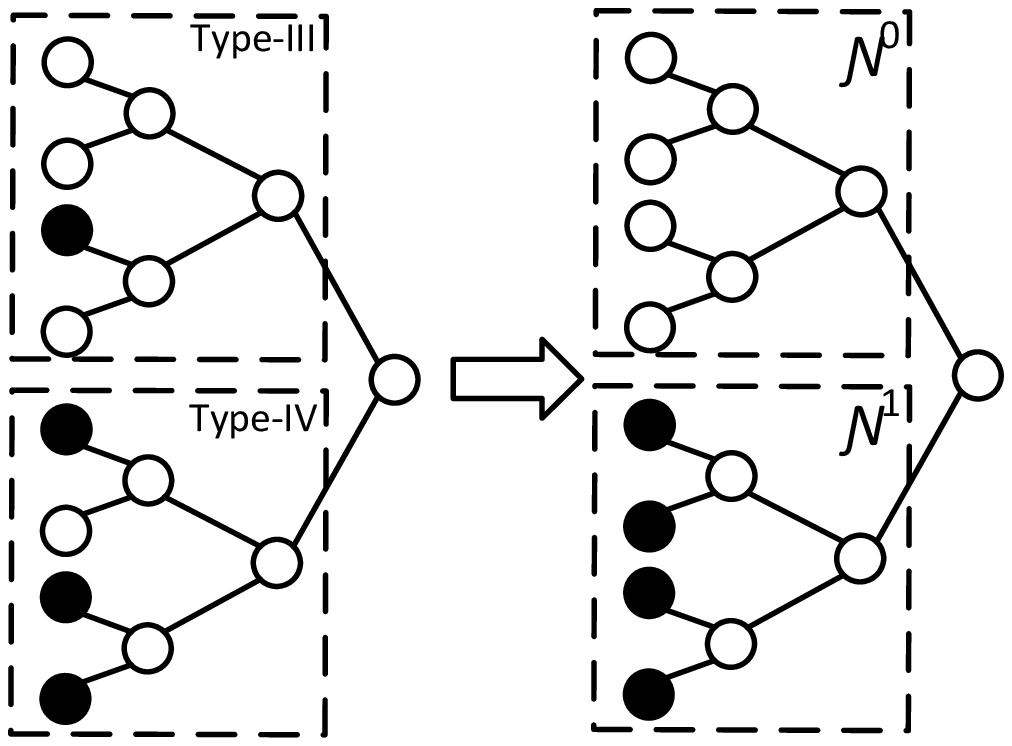}\label{switch}}
\caption{Examples of constituent codes division optimization 
%\protect\subref{n0n1} An example of $\mathcal{N}^0$ and $\mathcal{N}^1$ in a 8-bit %polar code tree, and \protect\subref{nsnr} An example of $\mathcal{N}^{SPC}$ and $%\mathcal{N}^{REP}$ in a 8-bit polar code tree
} 
\label{opt}
\end{figure}

According to T. Che's results~\cite{che2016tc}, there is only one clock cycle needed for decoding $\mathcal{N}^0$ and $\mathcal{N}^1$ node. 
Thus, it is unnecessary to optimize type-I and type-II codes since they are already fully optimized.
Our target should be focused on the type-III and type-IV. 
These two types are similar, they all only have one node with different type with others.
There are two situations we need to deal with.
The first situation is the optimization for single type-III or type-IV sub-tree. 
We can swap the different node with the first or last one to make it a $\mathcal{N}^{SPC}$ or $\mathcal{N}^{REP}$ nodes for type-IV or type-III sub-tree, respectively.
The second situation is that the optimization for a combination of type-III and type-IV sub-trees. 
For this case, we can swap the different node between the two types to make them became one type-I and one type-II sub-trees. 
For the first situation, suppose we have one Type-III node at stage $m+1$. 
It consist of one $\mathcal{N}^{REP}$ and one $\mathcal{N}^0$ node at stage $m$. 
This is shown in Fig~\ref{move}.
Totally, it needs $1+log_2{2^m}=m+1$ to finish decoding. If we move the place of the information bit to make it a $\mathcal{N}^{REP}$ constituent code, the new latency for decoding should be $log_2{2^{m+1}}=m+1$.
There is no change if we do this modification. This is also similar to type-IV situation.
For the second situation, suppose we have one type-III and one type-IV nodes at stage $m$ as shown in Fig.~\ref{switch}, the totally latency for decoding should be $2m+1$. If we swap the information bit in type-III and frozen bit in type-IV, we get one $\mathcal{N}^{0}$ and one $\mathcal{N}^1$ nodes. The total should be reduced to $2$. This makes a huge difference.
Thus, our target should be the swap operation between type-III and type-IV nodes.

Based on above discussion, the constituent code oriented polar code construction algorithm is proposed in algorithm~(1).
After we get the information and frozen bits positions using the traditional way, we apply this algorithm to adjust the location of some information and frozen bits to make more desired constituent code. This is an enhancement to the traditional construction method.  

\renewcommand{\algorithmicrequire}{ \textbf{Input:}} %Use Input in the format of Algorithm  
\renewcommand{\algorithmicensure}{ \textbf{Initialize:}} %UseOutput in the format of Algorithm  

\begin{algorithm}[t] 
\caption{constituent code oriented polar code construction}  
\begin{algorithmic}[1]  
\Require 
The set of Bhattacharyya parameter, $\bm{\epsilon_n}$; the initial set of bit property (frozen or information), $\bm{L}$; the threshold for bit swapping,$T_h$.
\Ensure $index = 1$. %算法的输出：Output  
\State get the sub-codeword-type look up table $T$ from $\bm{L}$; this table gives the sub-codeword type information and its index 
\If {$T[index]$ is type-III sub-codeword}
    \State get the index $i$ of the information bit in $T[index]$, search all the next type-IV sub-codewords and find the one whose frozen bit's Bhattacharyya parameter has the minimum difference with $\bm{\epsilon_n}[i]$. Recode the index $f$ of that. 
    \If{$\left|\bm{\epsilon_n}[i]-\bm{\epsilon_n}[f]\right|<T_h$}\State{swap the property of $\bm{L}[i]$ and $\bm{L}[f]$, update $T$. }\EndIf
\EndIf
\If {$T[index]$ is type-IV sub-codeword}
    \State get the index $f$ of the frozen bit in $T[index]$, search all the next type-III sub-codewords and find the one whose information bit's Bhattacharyya parameter is has the minimum difference with $\bm{\epsilon_n}[f]$. Recode the index $i$ of that. 
    \If{$\left|\bm{\epsilon_n}[i]-\bm{\epsilon_n}[f]\right|<T_h$}\State{swap the property of $\bm{L}[i]$ and $\bm{L}[f]$, update $T$.}\EndIf
\EndIf
\State increase $index$ by 1, repeat to $2$, until to the end of $T$ 
\end{algorithmic}
\end{algorithm}  

% \begin{figure}[!htc]
% \centering
% \subfloat[]{\includegraphics[width=1.5in]{observation.eps}\label{observation}}
% \hfil
% \subfloat[]{\includegraphics[width=1.5in]{diagonal_shift.eps}\label{diagonal_shift}}
% \caption{
% %SC decoding tree simplified by constituent codes
% \protect\subref{observation} Elements shift in generation matrix, and \protect\subref{diagonal_shift} diagonal cycle-shift in generation matrix 
% } 
% \label{shift}
% \end{figure}

% \subsection{Proposed Optimization Algorithm}
% \label{Proposed Optimization Algorithm}

\section{Simulation Results and Discussions}
\label{Simulation Results and Discussions}

\begin{figure}[!t]
\centering
\includegraphics[width=4in]{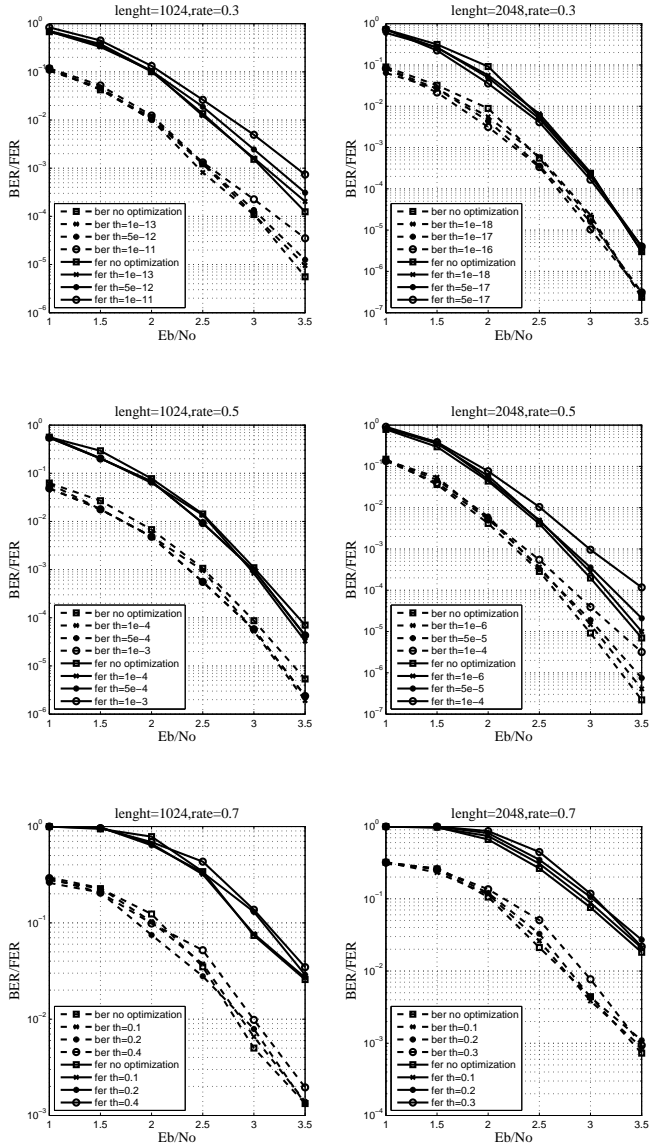}
\caption{The ber vs Eb/N0 performance for proposed construction scheme }
\label{performance}
\end{figure} 

Fig.~\ref{performance} shows the simulation results of proposed construction scheme. 
They are the BER/FER vs Eb/N0 performance for $1024$ and $2048$ length at different rate with multiple optimization thresholds. 
The polar codes is constructed based on the BEC channel and with target erasure rate $0.3$.
Table~\ref{latency_reduction} shows the decoding latency for each threshold  and the comparison with constituent codes decoding without any optimization and other state-of-the-art decoders. 
We even calculated the latencies for the codes with very long length such as $16384$.

According to Fig.~\ref{performance} and Table~\ref{latency_reduction}, we note that the proposed construction scheme generally is able to achieve at least around $20\%$ latency reduction with an negligible gain loss. 
For a certain code length, we can find that the acceptable threshold is increasing along with the code rate. 
This is due to the nature of channel polarization. 
There are more frozen and information bits mixed in the front and middle part of codeword for lower rate codes, which gives more flexibility during the optimization.
However, this does not indicate that this coding scheme is not working well on high rate. The interesting part is that the performance on high rate is as good as the low rate according to~Fig.~\ref{performance}.
It's possibly attributable to the following two reasons. First is that the coding performance of high rate itself is much worse than that of low rate. This causes the difference after optimization is not so obvious. 
The second reason is that the simulated code length is still not long enough due to the limitation of simulation environment. 
This is per coding theory that suggests the longer polar code will generally yield higher polarization.
%according to the polar coding theory, the longer the code, the more polarized they are. 
For a certain code rate, we can find the acceptable threshold is decreasing along with the code length. 
Since the longer codes are more polarized, the difference of each equivalent channel is becoming smaller and smaller as the length is increasing. 
This works fine with low and medium length but not so obvious with high length. This also can be explained by the two reason presented before. 

We compared the latency of proposed design with the decoder in~\cite{yuan2014low}; this is the fastest non-constituent-codes-based polar code decoder to the best of our knowledge. We can see even constituent codes decoder without any optimization is much faster than that. Our proposed construction scheme is capable of achieving $20\%$ or more latency reduction. This is very significant especially for very long code lengths. 

% It is noted that another approach that changing the frozen set to reduce the constituent codes related latency is reported in~\cite{giard2016fast}.
% Although the fundamental idea of our work and that in~\cite{giard2016fast} are similar, these two target two different decoder architectures and proposed two different constituent code oriented polar code construction methods. 
% Without prior access to~\cite{giard2016fast}, we independently propose our work in this paper.

Compared with~\cite{giard2016fast} in which also changes the frozen and information sets to benefit the decoding, our work has two main differences. The first one is that we target a different decoder architecture, which results in different demanding of desirable constituent codes combination. The second difference is that the construction algorithm is different.      

\begin{table}[t]
\centering
\caption{Latency reduction}
\label{latency_reduction}
\scalebox{0.85}[0.85]{
\begin{tabular}{|c|c|c|c|c|c|}
\hline
decoder                    & length                 & rate                  & threshold       & latency & reduction(\%)      \\ \hline
\multirow{36}{*}{proposed} & \multirow{12}{*}{1024} & \multirow{4}{*}{0.3}  & no optimization & 303     &                    \\ \cline{4-6} 
                           &                        &                       & 1e-13           & 288     & 4.9                \\ \cline{4-6} 
                           &                        &                       & 1e-12           & 260     & 14                 \\ \cline{4-6} 
                           &                        &                       & 1e-11           & 234     & 22.7               \\ \cline{3-6} 
                           &                        & \multirow{4}{*}{0.5}  & no optimization & 266     & -                  \\ \cline{4-6} 
                           &                        &                       & 1e-4            & 255     & 4.1                \\ \cline{4-6} 
                           &                        &                       & 5e-4            & 218     & 18                 \\ \cline{4-6} 
                           &                        &                       & 1e-3            & 197     & 21.8               \\ \cline{3-6} 
                           &                        & \multirow{4}{*}{0.7}  & no optimization & 172     & -                  \\ \cline{4-6} 
                           &                        &                       & 0.1             & 165     & 4                  \\ \cline{4-6} 
                           &                        &                       & 0.2             & 137     & 20                 \\ \cline{4-6} 
                           &                        &                       & 0.4             & 126     & 23.6               \\ \cline{2-6} 
                           & \multirow{12}{*}{2048} & \multirow{4}{*}{0.3}  & no optimization & 576     & -                  \\ \cline{4-6} 
                           &                        &                       & 1e-18           & 549     & 4.6                \\ \cline{4-6} 
                           &                        &                       & 1e-17           & 519     & 9.9                \\ \cline{4-6} 
                           &                        &                       & 1e-16           & 493     & 14.4               \\ \cline{3-6} 
                           &                        & \multirow{4}{*}{0.5}  & no optimization & 493     & -                  \\ \cline{4-6} 
                           &                        &                       & 1e-6            & 487     & 1.2                \\ \cline{4-6} 
                           &                        &                       & 1e-5            & 436     & 10.5               \\ \cline{4-6} 
                           &                        &                       & 1e-4            & 323     & 33.6               \\ \cline{3-6} 
                           &                        & \multirow{4}{*}{0.7}  & no optimization & 297     & -                  \\ \cline{4-6} 
                           &                        &                       & 0.1             & 269     & 9.4                \\ \cline{4-6} 
                           &                        &                       & 0.2             & 248     & 16.5               \\ \cline{4-6} 
                           &                        &                       & 0.3             & 228     & 23.2               \\ \cline{2-6} 
                           & \multirow{12}{*}{16384}  & \multirow{4}{*}{0.3}  & no optimization & 3992  & -                  \\ \cline{4-6} 
                           &                        &                       & 1e-50           & 3661    & 8.3                \\ \cline{4-6} 
                           &                        &                       & 1e-45           & 3242    & 18.8               \\ \cline{4-6} 
                           &                        &                       & 1e-40           & 2721    & 31.8               \\ \cline{3-6} 
                           &                        & \multirow{4}{*}{0.5}  & no optimization & 3327    & -                  \\ \cline{4-6} 
                           &                        &                       & 1e-13           & 3187    & 4.2                \\ \cline{4-6} 
                           &                        &                       & 1e-12           & 2898    & 12.9               \\ \cline{4-6} 
                           &                        &                       & 1e-11           & 2465    & 25.9               \\ \cline{3-6} 
                           &                        & \multirow{4}{*}{0.7}  & no optimization & 1350    & -                  \\ \cline{4-6} 
                           &                        &                       & 0.1             & 1260    & 6.6                \\ \cline{4-6} 
                           &                        &                       & 0.2             & 1165    & 13.7               \\ \cline{4-6} 
                           &                        &                       & 0.4             & 898     & 33.4               \\ \hline
\multirow{3}{*}{\cite{yuan2014low}}      & 1024                   & \multicolumn{2}{c|}{\multirow{3}{*}{-}} & 767     & \multirow{3}{*}{-} \\ \cline{2-2} \cline{5-5}
                           & 2048                   & \multicolumn{2}{c|}{}                   & 1535    &                    \\ \cline{2-2} \cline{5-5}
                           & 16384                  & \multicolumn{2}{c|}{}                   & 12287   &                    \\ \hline
\end{tabular}
}
\end{table}

\section{Conclusion}
\label{Conclusion}

This paper presented a novel polar code construction scheme which reduces the decoding latency.
The proposed constituent codes oriented polar code construction algorithm can automatically produce more types of constituent codes which are desirable.  
The simulation results show that the proposed construction scheme generally is able to achieve at least around $20\%$ latency deduction with negligible decoding performance loss. 
Besides, compared with non-constituent-codes-based decoder, the constituent codes based decoder has a measurable advantage in term of latency. Our construction scheme is able to further enhance the timing performance. 

\bibliographystyle{IEEEtran}
\bibliography{IEEEabrv}

\end{document}